# Privacy and Security in IPv6

Dr. Nour El-Kadri, Sowmyan Jegatheesan

University of Ottawa, Ontario, Canada
jsowmyan@gmail.com

**Abstract**

*Many Internet service providers (ISPs) throughout the world are now in the process of integrating IPv6 into their Internet access products for retail customers and corporate clients. One of the most important features of the IPv6 protocol is its huge address space. This will enable users to assign a life-long stable, unique and globally routable IP address to each personal device. All users will be able to set up public services at home and be able to communicate from and to their devices from any point in the network. The resulting end-to-end reachability is considered a major improvement over the current situation [1]. IPv6 evoked a new debate between privacy and security in order to attract more unique IP addresses to locate users, especially with the advent of mobile devices. While the initial IPv6 specification from 1995 does already contain a variety of security considerations (namely, message integrity protection as well as encryption based on IPsec), privacy aspects have not played a major role in the beginning. [2]*

## 1. Introduction

This work is mainly about the privacy issues and its consequences in the area of security. There are increasing concerns about what the corporate world is focusing at and what the regulators are going to do to make things perfectly fine in protecting the individual rights. Various issues that would be in our focus will include

- Auto-Configured addresses.
- IPv6 Privacy Addressing
- Security in IPv6
- Complications in Privacy

## 2. Auto-Configured addresses

Auto configured addresses, the default addressing system in IPv6, provides a third party a means to track and monitor targeted users globally using simple tools such as ping and trace route. Signed messages also expose the identities of both the sender and receiver to a third party. [3]

SLAAC which is called as Stateless Address Auto configuration is a method IPv6 hosts use to self-configure addresses. This was aimed at reducing the management burdens of the network administrators, and enabling the hosts to self or auto configure. This seemed to be a remarkable change from IPv4 because of the elimination of DHCP servers which were used to issue addresses to the hosts. Having the hosts configure their own addresses would reduce the workload of the network administrators.

The host addresses or interface identifiers (IIDs) stay the same regardless of the subnet they connect to in the SLAAC method. The default addressing scheme, referred to as the 64-bit extended unique identifier (EUI-64), uses the MAC address as the IID. [4] This would result in a huge expose to the attacker who would know the list of subnets and the host MAC addresses. This would also mean that the addresses and hosts are vulnerable to tracking and also easy targets for attacks from anywhere around the world.

Privacy extensions are important when it comes to securing the client, as they can only protect the client from being attacked and the question about the security of the server is still unanswered. Also privacy extensions do not change often enough to prevent network attacks; they are less effective for globally available systems that require static addressing to ensure connectivity, like web servers or VPN endpoints. Privacy extensions are primarily implemented for web traffic communications. Applications, such as VoIP and VPNs, cannot function with privacy extensions. Also systems with Privacy Extensions are also easy targets for attackers.

The privacy extensions used by the Windows OS relies on another IPv6 address that is used in neighbor discovery, local DNS, and other functions. This address is static and is reachable by other hosts. A target machine can be attacked by an attacker who observes this address.

## 3. IPv6 Privacy Addressing

Each IPv6 address includes the MAC address of a specific computer and not of the router. We can avoid exposing MAC addresses by using IPv6 Privacy Addressing. When enabled, the system will generate a temporary address with a random suffix in addition to the EUI-64-based address. The following is the procedure to be followed to enable the same. [5]

**Windows OS**
Use the following in the command line
- netsh interface ipv6 set privacy state=enabled

**Linux**
To enable temporary addresses and make them preferred for outgoing connections:
- sysctl net.ipv6.conf.all.use_tempaddr=2
- sysctl net.ipv6.conf.default.use_tempaddr=2

To enable temporary address generation, but keep the old (Autoconf) address as preferred:
- sysctl net.ipv6.conf.all.use_tempaddr=1
- sysctl net.ipv6.conf.default.use_tempaddr=1

**Mac OS X**
- Enabled by default since OS X 10.7 Lion

If the hardware address is used in the IPv6 address, it means the network uses IPv6 Stateless Auto configuration. We can simply pick our own address suffix and configure IPv6 manually. Though the manually added address will not have the hardware info, it will still be static unlike Privacy Addressing, which changes addresses very often. Also, static addresses can become very complex in a network larger than 2-3 devices.

## 4. Security in IPv6

Patrons of IPv6 argue that the Internet protocol Version 6 ensures that the addresses are difficult for attackers to attack or they are not as vulnerable as projected. They compliment this by saying the increased number of IP addresses would bring in about new conflicts between privacy and security. The uniqueness of IPv6 globally would cause security concerns with many users as they would be easily tracked down and this is a clear breach of privacy as there is visibility on the devices using the newer version.

In IPv6, each subnet now has 18 quintillion addresses, instead of 256 addresses per subnet. It will now take longer for cybercriminals to find the targets and it also eliminates one of the primary vectors of malware spread. Common cyber attacks such as social engineering, Trojans, worms, and viruses will continue to work on upper layers and not the protocol itself as long as there are people who would fall victims for the tricks of the attackers!

Experts feel that if users tried to use encryption on IP addresses to protect their privacy, it would complicate network security management in terms of intrusion prevention system (IPS) such that network inspection and protection is compromised. [6] Some legitimate users would also be prevented from accessing sites because of the encrypted notion of their IP address. This is already an issue in the IPS that tries to minimize false positives. Individuals and organizations are free to choose what technology they want use based on the technologies available to them and to the scale of security and privacy, considering the configuration for their need.

## 5. Complications in Privacy

There are many complications when it comes to IPv6. Many technical experts, Internet Gurus and Eminent scientists have voiced their concerns over this. IPv6 addresses are unique to each device and are almost abundantly available for use and the volume issue that has crippled IPv4 is resolved here. The problem with IPv6 arises on this particular issue. When there is direct relationship with the device and the access number Individually Identifiable Information (III) is easily retrievable.

The marketing industry and experts targeting specific users for getting the products to our own devices would be overjoyed about this particular aspect as they would always know from which device what was searched for or what content was accessed by which IPv6 address and the device identifier associated with it. This will surely make the advertisers more eager to push for and pressure the advertisement servers to give targeted and personalized results and if there are no specific regulations then the information about the users, their personal preferences and personal information would be available and known to all.

There are always people who tend to bend the laws but without any ethical protocols, laws behind this or even rulings or regulations it would be foolish from our side to trust, believe and understand that Individually identifiable information would not be compromised.

The problem is, it's very complicated because there will have to be some way to regulate and verify obfuscation or hashing of III and unique IPv6 ID's [7] .What was dubbed 'World IPv6 Day,' on June 8, 2012 there were tests by Google, Yahoo!, amongst others, to see if they would incur issues similar to this before switching over to calling IPv6 addresses only. The switch is imminent and there arises a question on what regulations are there to prevent the use of PII by the advertising and digital marketing community.

There is widespread speculation that the Advertising community and the advertising platforms are approaching the federal government or the congress to make things easier for them and to make the best use of Individually Identifiable Information. The corporate lobbying would prevent Washington to take strong steps until unless people are aware about this and start protesting like the Wall Street protests or the Copyright infringement protests carried out by with the support of Wikipedia, Google etc

## 6. Conclusion

There are newer methods of IPv6 prefix alteration schemes that would help to improve the protection of residential Internet connections. The introduction of IPv6 is very less opportunity to improve the overall level of privacy for many users. IPv6 prefix alteration and similar techniques cannot be a permanent privacy solution for all applications on the Internet. As long as technologies such as HTTP cookies are enabled on higher protocol levels, tracking is obviously still possible. [8]